\documentclass[20pt,APS,balancelastpage,nofootinbib,twocolumn,fleqn,showkeys]{revtex4}
\usepackage{amsfonts}
\usepackage{amsmath}
\usepackage{amssymb}
\usepackage{color}
\usepackage{float}
\usepackage{hyperref}
\usepackage{subfigure}
\usepackage[rightcaption]{sidecap}
\usepackage{graphicx}
\begin{document}
\title{Complex dynamics of nano-mechanical membrane in cavity optomechanics}
\author{Muhammad Javed Akram}
\email{javed\_quaidian@yahoo.com}
\author{Farhan Saif}
\email{farhan.saif@fulbrightmail.org}
\affiliation{Department of Electronics, Quaid-i-Azam University, 45320  Islamabad, Pakistan.}
\begin{abstract}
Theoretical analysis of a suspended nanomechanical membrane subject to an optical driving field in an optomechanical cavity is presented, which is
confirmed through numerical simulations. In the presence of an optical field between its mirrors, the high-finesse optomechanical resonator acts as an oscillator driven by a radiation pressure force. The periodic nature of the radiation pressure force makes the nano-mechanical membrane in the optomechanical system as a kicked harmonic oscillator. Mathematically the physical system displays a stochastic web map that helps to understand several properties of the kicked membrane in classical phase space. We find that our web map is area preserving and displays quasiperiodic symmetrical structures in phase space which we express as $q$-fold symmetry. It is shown that under appropriate control of certain parameters, namely the frequency ratio and the kicking strength, the dynamics of kicked membrane exhibits chaotic dynamics. We provide the stability analysis by means of Lyapunov exponent and survival probability.
\end{abstract}
\maketitle

\section{Introduction}
\label{intro}
Mechanical oscillators coupled to optical degrees of freedom in a nano-cavity via radiation pressure have received immense attention as an important frontier in the field of quantum optics around the globe \cite{1,2,3,4,5,6,7,8}. Recent experimental advances have led to observe mechanical dynamics provided by radiation pressure in the laboratory. Owning the discrete nature of photons, the quantum fluctuations of the radiation pressure give rise to the standard quantum limit \cite{9,10,11}. In addition to the \textit{quantum back-action} effect, the radiation pressure of light confined within a resonator, give rise to the effect of dynamical back-action \cite{12}, which is a \textit{classical effect} caused by finite cavity decay time. These devices have been suggested to analyze the optical bistability \cite{13} and nonlinear dynamics \cite{14,15}. On the other hand, rapid developments in atom optics \cite{16} have made this subject a testing ground for the dynamical localization~\cite{17,18,19,20,21}, dynamical recurrences \cite{22,23,24}, bifurcation response \cite{25}, and classical quantum transition \cite{26,27,28}.

In classical mechanics, the extreme sensitivity to initial conditions becomes the hallmark of chaos \cite{29,30,31,32,33,34}. In addition, it is well known that Hamiltonian systems are carriers of chaos, and in general, the phase space of such a system contains regions where regular motion is accompanied by stochasticity \cite{35,36,37,38}. Certain subtle properties of chaos can be linked to the phenomenon of the diffusion of particles and to the problem of the symmetry of the tilings of a space \cite{39}. The existence of stochastic webs is an important physical phenomenon since infinite particle transport is performed along the channels of the web. Hence, the skeleton of the web tiles the phase space and imposes a dynamical origin on symmetry groups (namely, the
crystalline and quasicrystalline symmetry). The kicked harmonic oscillator is an example of such a system \cite{39,40,41,42,43,44}. In the present work, we model the optomechanical system in which nano-mechanical membrane experiences periodic kicks. We report that the classical system exhibits complex evolution in phase space. As a result, we obtain crystalline as well as quasicrystalline symmetrical structures. Our contribution leads to investigate a complex symmetry in optomechanical systems by developing a web map using methods of Hamiltonian dynamics.

The paper is developed as follows: In Sect.~\ref{sec:2}, we present the experimental model of the system.We introduce the dynamical mapping in Sect.~\ref{sec:3}. Several properties of the kicked membrane have been explored in Sect.~\ref{sec:4}. In Sect.~\ref{sec:5}, we investigate various aspects of stability and unstability of the system by means of Lyapunov exponent and survival probability. Finally, the results are summarized in Sect.~\ref{sec:6}.
\section{The Model}\label{sec:2}
\begin{figure}[t]
\includegraphics[width=0.45\textwidth]{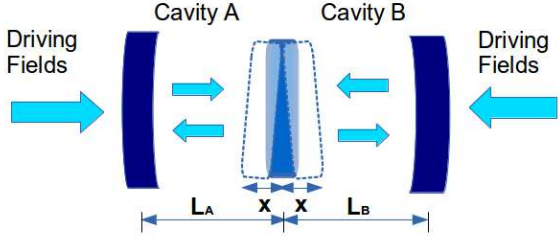}
\caption{The schematics of the optomechanical system: a movable nano mechanical membrane with two perfectly reflecting surfaces is suspended inside a driven cavity with two transmissive fixed mirrors.} \label{cavity}
\end{figure}

Our nano-optomechanical system consists of a Febry-P\'erot cavity, which is coupled with two fixed end-mirrors, as shown in Fig.~\ref{cavity}. A thin mechanical membrane with perfectly reflecting surfaces is placed inside the cavity. Similar system has been studied to cool mechanical membrane \cite{45}. 
The Hamiltonian of system can be written as
\begin{equation}
 H =H_c + H_m,
\end{equation}
where $H_m=p^2/2m + m\omega^2x^2/2$ is the Hamiltonian for the membrane with effective mass $m$ and the bare eigen-frequency $\omega$, and
  \begin{eqnarray}
H_c = \hbar(\omega_A - G_Ax)a^\dag a + \hbar(\omega_B + G_Bx)b^\dag b \nonumber \\
+\hbar(\xi_A e^{iv_At}a + \xi_Be^{iv_Bt}b + H.C.),
  \end{eqnarray}
describes the optomechanical coupling between single-mode optically driven cavities A and B respectively with $a(a^\dagger)$ and $b(b^\dagger)$ being lowering (raising) field operators. Here, $\omega_{A,B}$ and $\omega$ respectively, are the field frequencies for cavity $A$ and $B$ and the membrane. Moreover, $\xi_{A,B}$  are the amplitudes of external optical driving fields to the cavities with corresponding driving frequencies $\nu_{A,B}$, $G_A=\omega_A/L_A$ and $G_B=\omega_B/L_B$ are the corresponding optomechanical coupling strengths via the radiation pressure, and $L_{A,B}$ are the lengths of cavities A and B.

Under the same condition as in \cite{45}, we obtain the effective Hamiltonian,
\begin{equation}
     H_{eff}=\dfrac{p^2}{2m} + \dfrac{1}{2}m\omega^2 x^2 -  \dfrac{2\hbar\xi^2}{\Delta}(1 + \dfrac{G^2}{\Delta^2}x^2).
\end{equation} \label{Ham}
The symbol $\xi$ is the amplitude of external optical driving fields with corresponding driving frequencies $\nu$. Here, $G=\omega_A/L$ is the corresponding optomechanical coupling strength via radiation pressure and $\Delta=\omega_A-\nu$ is the optical detuning. Note that, the last term in the above Hamiltonian can be approximated to $\cosh\alpha x$, under the condition $x << \lvert \dfrac{\Delta}{\omega_A}\rvert L_A$, where $\alpha=\dfrac{\sqrt{2}\omega_A}{L\Delta}$. The effective Hamiltonian becomes,
\begin{equation}
     H_{eff}=\dfrac{p^2}{2m} + \dfrac{1}{2}m\omega^2x^2 - k\cosh\alpha x.
\end{equation}
Here $k=2\xi^2/\Delta$. We also ignore the effects from cavity decay rates $\kappa_A=\kappa_B=\kappa$ of cavity fields based on the fact that $\kappa$ is much less than the detuning \cite{2,45}. Moreover, as we consider the short time interaction between optical field and membrane into account, this periodic nature of the radition pressure force makes the nano-mechanical membrane in the optomechanical system as kicked harmonic oscillator.

We may express the periodic kicks by $\xi=\xi_o\sum_{n=- \infty}^{n= \infty}\delta(t-nT)$ which yields the effective Hamiltonian as,
 \begin{equation}
     H_{eff}=\dfrac{p^2}{2m} + \dfrac{1}{2}m\omega^2x^2 - k\cosh\alpha x\sum_{n=-\infty}^{n=\infty}\delta(t-nT),
\end{equation}
in which perturbation is a periodic sequence of $\delta$-pulses (i.e. kicks) following with the time period $T=\dfrac{2\pi}{\nu}$ and $k$ ($k=\dfrac{2\xi_o^2}{\Delta}$) correspond to amplitude of the pulses. Note that in the absence of kicks i.e. for $k=0$ (when there are no external fields), the motion of the membrane is simple harmonic. Interestingly, the effective Hamiltonian that we have obtained so far, describes the dynamics of the kicked membrane just like a conventional $\delta$-kick harmonic oscillator. Since, in the presence of $delta$-kicks, the dynamics of the oscillating membrane switches from harmonic to forced, and we may express it by the following classical Hamilton equations,
\begin{eqnarray}
&\dot{p}& = -m\omega^2 x+k\alpha \sinh \alpha x\sum_{n=- \infty}^{n= \infty}\delta(t-nT) \label {momentum} \\
&\dot{x}&  = \frac{p}{m} \\
&\ddot x& + \omega^2 x = \frac{-k\alpha \sinh x}{m}\sum_{n=- \infty}^{n= \infty}\delta(t-nT)
\label{motion}
\end{eqnarray}
\section{Dynamical Mapping: The Stochastic Web Map}\label{sec:3}
It is obvious that between two consecutive interactions i.e. kicks, the time evolution is simply that of a conventional harmonic oscillator, given by \cite{46}
 \begin{eqnarray}
x(t)&=& x(0)\cos(\omega t) + \dfrac{p(0)}{m \omega} \sin \omega t, \nonumber\\
p(t)&=& p(0)\cos(\omega t) - m \omega x(0)\sin \omega t. \label{sol}
\end{eqnarray}
In order to understand the evolution at the time of kick, we integrate Eq.~(\ref{momentum}), at an arbitrary kick $nT$. This yields
\begin{eqnarray}
p(nT&+&\epsilon)-p(nT-\epsilon)=-m\omega^2 x + \int_{nT-\epsilon}^{nT+\epsilon} x(t)\,dt \nonumber\\
&+&k\alpha \int_{nT-\epsilon}^{nT+\epsilon} \sinh\alpha x \sum_{n=- \infty}^{n= \infty}\delta(t-nT) \,dt,
\end{eqnarray}
where $\epsilon$ tends to zero. The equation connects the momentum just before $n$th impact with momentum just after the kick.  As $\epsilon$ tends to zero the remaining integral in above equation vanishes, leaving us with
\begin{equation}
p(nT^+)=p(nT^-)+k\alpha \sinh{\alpha x(nT)}.
\end{equation}
\begin{figure*}[t]
\centering
{\includegraphics[width=0.32\textwidth]{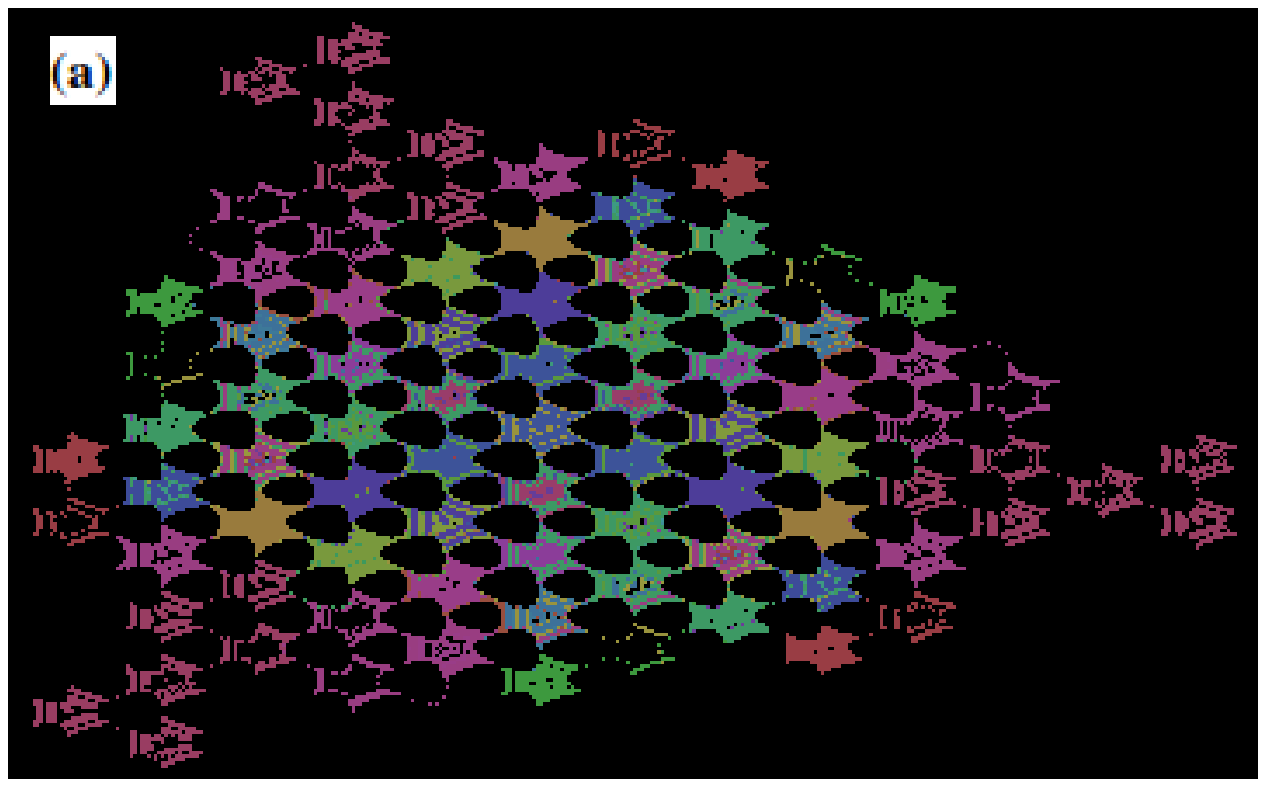}}\label{q3}
{\includegraphics[width=0.32\textwidth]{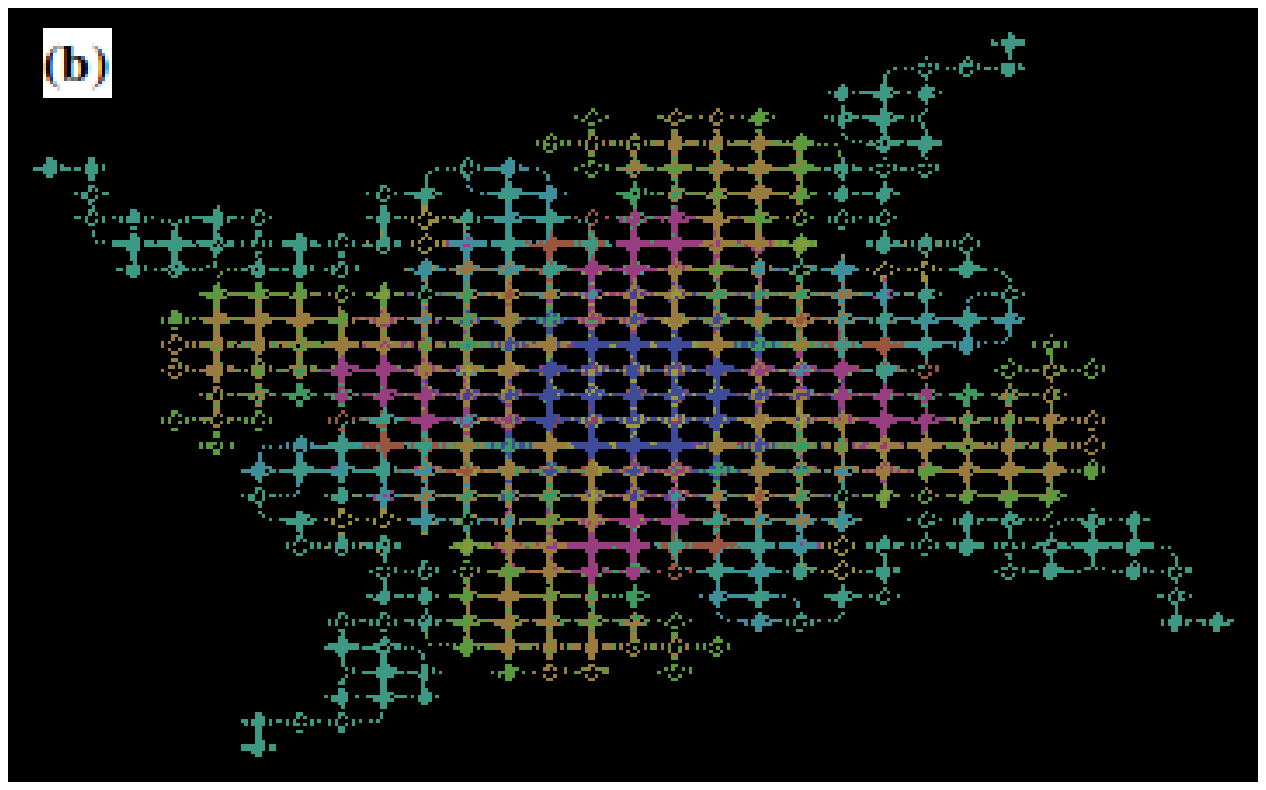}}\label{q4}
{\includegraphics[width=0.32\textwidth]{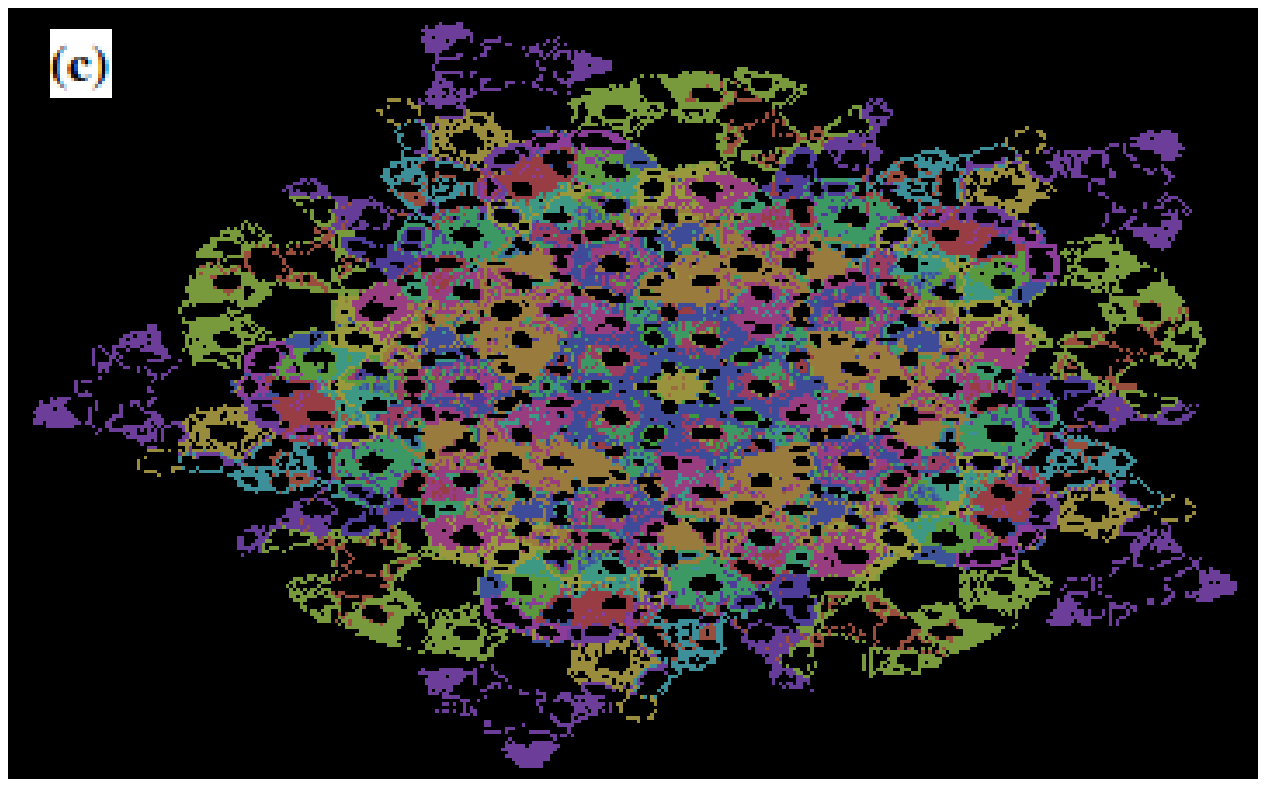}}
{\includegraphics[width=0.32\textwidth]{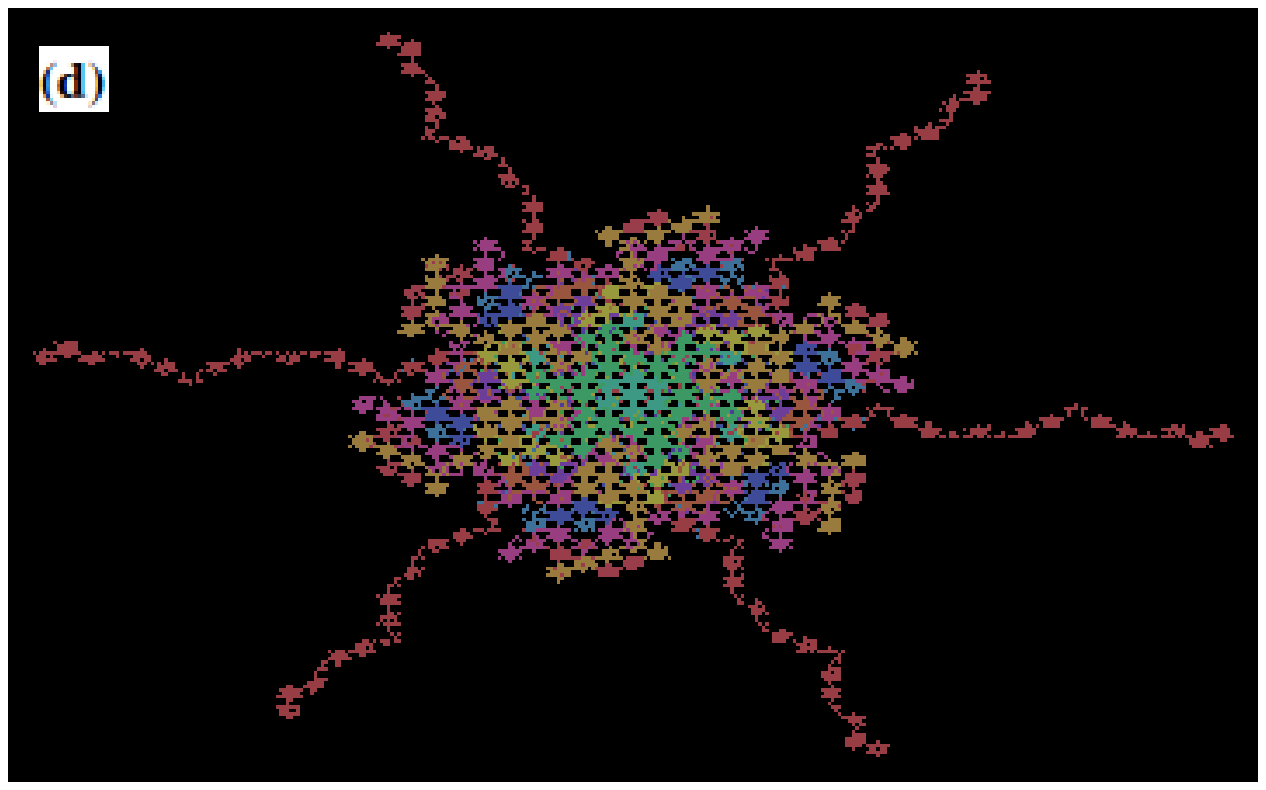}}
{\includegraphics[width=0.32\textwidth]{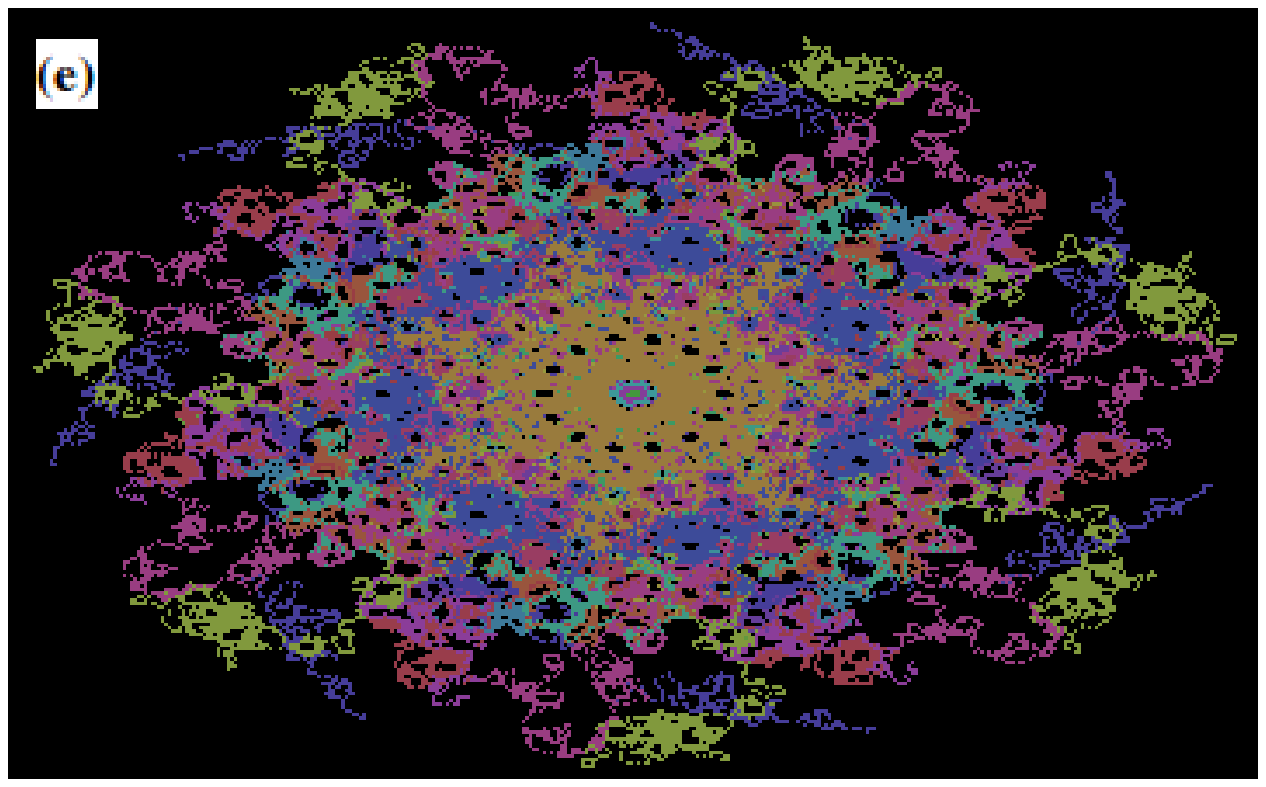}}
{\includegraphics[width=0.32\textwidth]{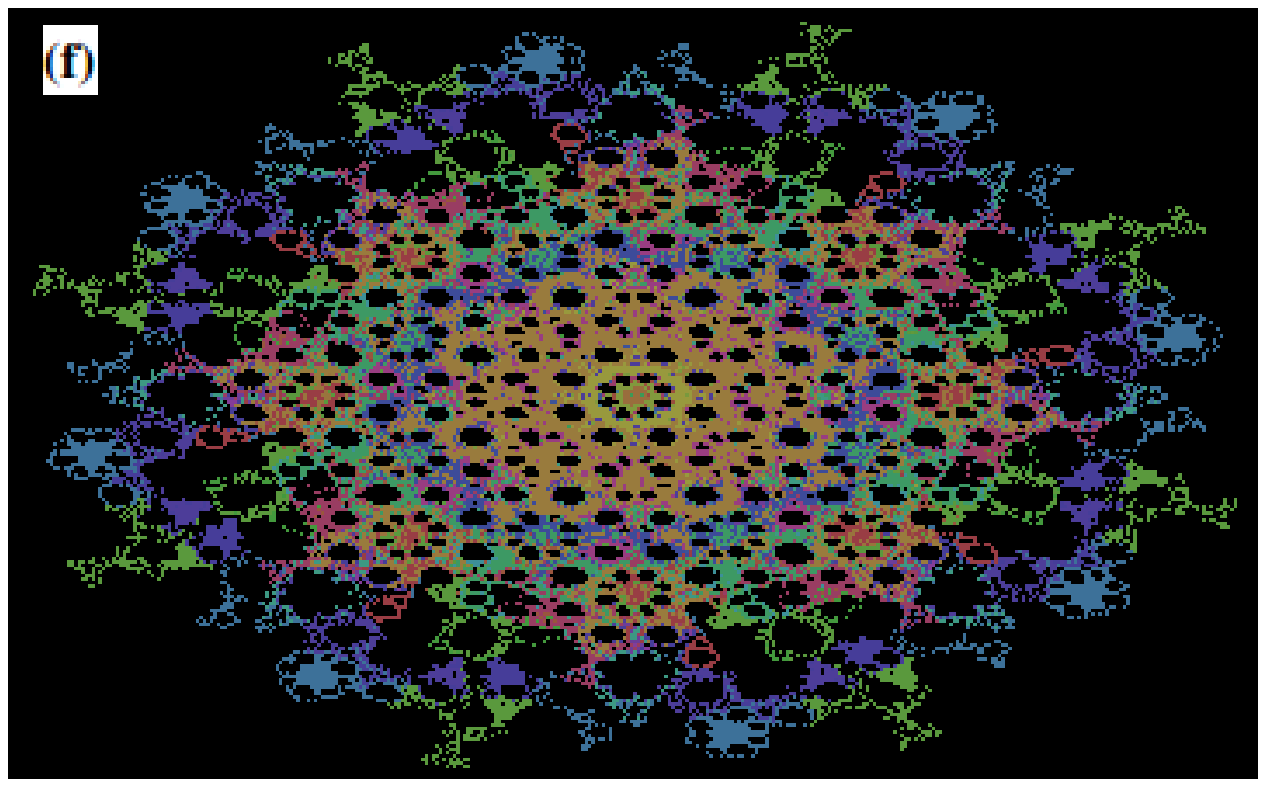}}
\caption{Phase portraits displaying $q$-fold symmetry for {\bf a} $q$ = 3,
{\bf b} 4, {\bf c} 5, {\bf d} 6, {\bf e} 7 and {\bf f} 8, where kicking strength is $K= 0.01$ for all the cases. Scaled position is plotted along horizontal axis, whereas the vertical axis indicates scaled momentum coordinate.
Parameters of the optomechanical system are: $L = 2$ mm, $\Delta/2\pi = 107$ Hz, $m = 50$ Pg, $\omega/2\pi = 134$ KHz and $\omega_{A}/2\pi = 7\times 10^{14}$ Hz}
\label{phase portraits}
\end{figure*}
We can now combine this with the equations of motion for the harmonic oscillator to derive a mapping from just before one kick to just before the next. We introduce the convenient shorthand $x_n$ and $p_n$.
\begin{eqnarray}
x_{n+1}&=& ax_n + \dfrac{1}{m\omega}(p_n + k \sinh \alpha x_n)b \nonumber\\
p_{n+1}&=& -m \omega b x_n + a(p_n + k \sinh \alpha x_n)\label{map1}
\end{eqnarray}
where $a=\cos \omega t$ and $b=\sin \omega t$. Before we start exploring the dynamics of the web map obtained in Eq.~(\ref{map1}), a better choice is re-scaling the variables $x$ and $p$ to dimensionless quantities in order to reduce the number of free parameters, by defining $\bar{x}=\alpha x$ and  $\bar{p}=\dfrac{\alpha p}{m\omega}$. We also introduce $\bar{t}=\omega t$ and  $\bar{T}=\nu T$. Thus, kick to kick mapping with dimensionless parameters can be rewritten as:
\begin{eqnarray}
\bar{x}_{n+1}&=& a\bar{x}_n + (\bar{p}_n + K \sinh \bar{x}_n)b \nonumber\\
\bar{p}_{n+1}&=& (\bar{p}_n + K \sinh \bar{x}_n)a\ - b \bar{x}_n. \label{web-map}
\end{eqnarray}
where, $K=\dfrac{\alpha k}{m\omega}$. It is now apparent that we have exactly two free parameters: the dimensionless kicking strength $K$, and the dimensionless frequency ratios $q$ (or kick to kick period) given by,
\begin{equation}
\alpha_q=\dfrac{2\pi}{q}.\label{resonance}
\end{equation}
In the next section we see that the parameter $\alpha_q$ plays an important role in the evolution. For simplicity we omit bar from the variables i.e. $\bar{x}=\alpha x$ and  $\bar{p}=\dfrac{\alpha p}{m\omega}$ in later discussions. In the next section we study the effect of parameters $K$ and $q$ on the phase space structures and $q$-fold symmetry.
\section{Dynamical Characteristics}\label{sec:4}
It should be noted that for kicked systems there occur two frequencies: one natural frequency of the system and the frequency of the periodic kicks. So, the dynamics produced by the mapping described in Eq.~(\ref{web-map}) can be grouped into two broad categories: one where the frequency ratio $q$ is a rational number, that is the membrane is kicked a rational number of times per harmonic oscillator period; and the second, where $q$ is an irrational number. We concentrate on the case where $q$ is rational, which has interesting particular properties showing very fascinating patterns as a phase portrait. Sample of these stroboscopic Poincar\'e sections of the dynamics described by the mapping of Eq.~(\ref{web-map}) are shown in Fig.~\ref{phase portraits}.

The initial condition in each of the cases displayed in Fig.~\ref{phase portraits} is iterated over $15,000$ kicks. In order to explicitly demonstrate the complex dynamics of nano- mechanical membrane, we use the experimental parameters from Refs.~\cite{47,48} such that: the length of the optomechanical cavity $L=2$ mm, the optical detuning $\Delta/2\pi=10^7$ Hz, the effective mass of the membrane $m=50$ Pg, the bare eigenfrequency of the membrane $\omega/2\pi=134$ KHz, the
eigenfrequency of the optical cavity $\omega_A/2\pi=7 \times 10^{14}$ Hz. These parameters make it possible to realize theoretical work in present day laboratory experiments. In numerical experiments, we take the value of $K=0.01$ whereas the frequency ratio $q$ governs the $q$-fold symmetry i.e. in Fig.~\ref{phase portraits} we show phase portraits displaying: (a) \textit{three-fold symmetry}, (b) \textit{four-fold symmetry} and so on. We see that there is obviously a high degree of rotational symmetry as well as self similarity in the trajectories taken. The symmetry is essentially rotational in phase space. However, the unpredictable part of the point particle's dynamics moving in phase space is more in knowing how it moves in and out through phase space. The symmetric structure brought out by these stable and unstable dynamics makes \textit{stochastic web}, as there is an interconnected web of channels which spread through all of the phase space.

Note that, the dynamically obtained $q$-fold symmetry for the resonance condition in Eq.~(\ref{resonance}), is of the crystalline type for $q\in\lbrace3,4,6 \rbrace$ and the quasi-crystalline type for $q\in\lbrace5,7,8\rbrace$. In all phase portraits in Fig.~\ref{phase portraits}, the dots belong to a trajectory in the domain of chaotic dynamics (stochastic sea) and the curves are closures at the Poincar\'e map for quasi-periodic structures, also known as islands of stability \cite{39}. Isolated domains in chaotic motion exists inside the islands. Interestingly, a magnification of an island shows resemblance with main pictures in Fig.~\ref{phase portraits}. Thus, the self-similar behaviour confirms the complexity of the chaotic dynamics and existence of fractals.
The coexistence of both regions of stable dynamics and chaos in the phase space, enables us to analyze the onset of chaos and the appearance of the minimal region of chaos \cite{39}. Moreover, in contrast to the classification of global phase portraits of planar quartic quasi-homogeneous polynomial differential systems \cite{49} and bounded solutions of a complex permanent magnet synchronous motor (PMSM) system \cite{50}, these phase portraits contains stable as well as complex quasi periodic structures.

\begin{figure}[ht]
\centering
{\includegraphics[width=0.2\textwidth]{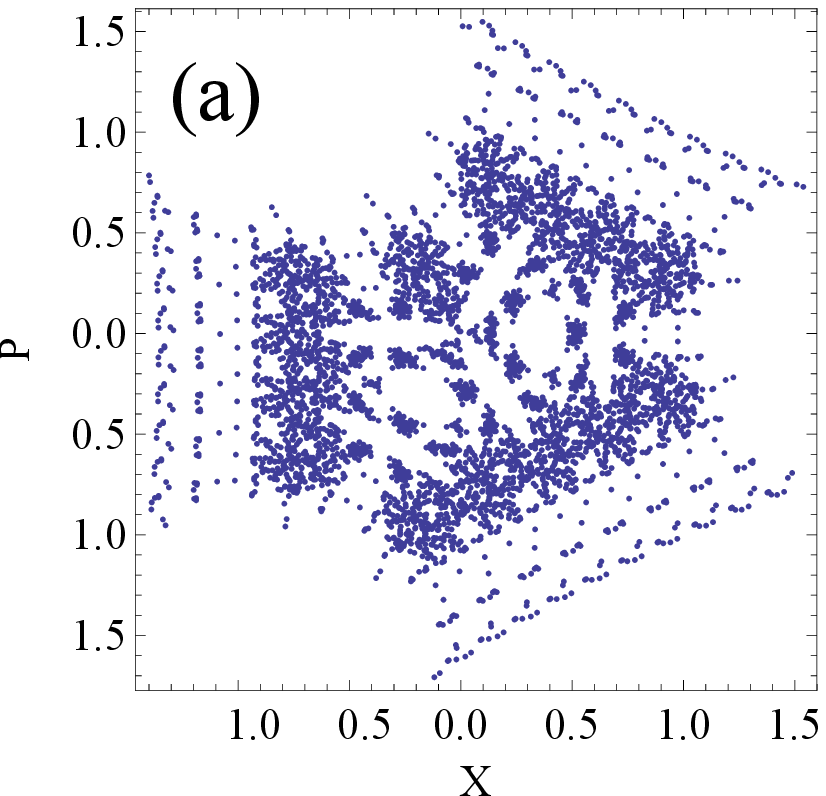}}
{\includegraphics[width=0.2\textwidth]{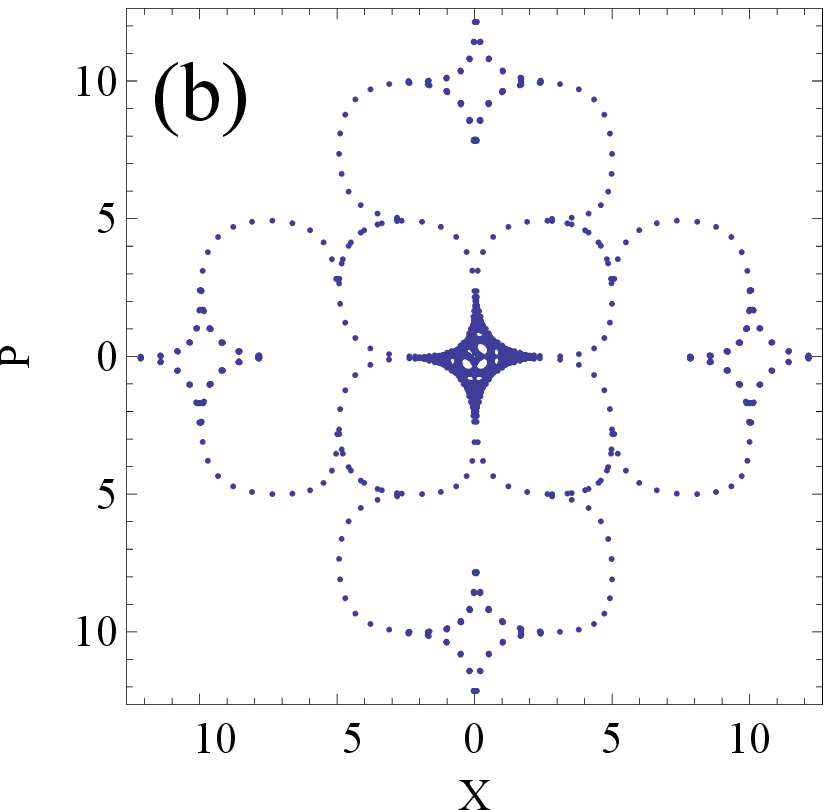}}
{\includegraphics[width=0.2\textwidth]{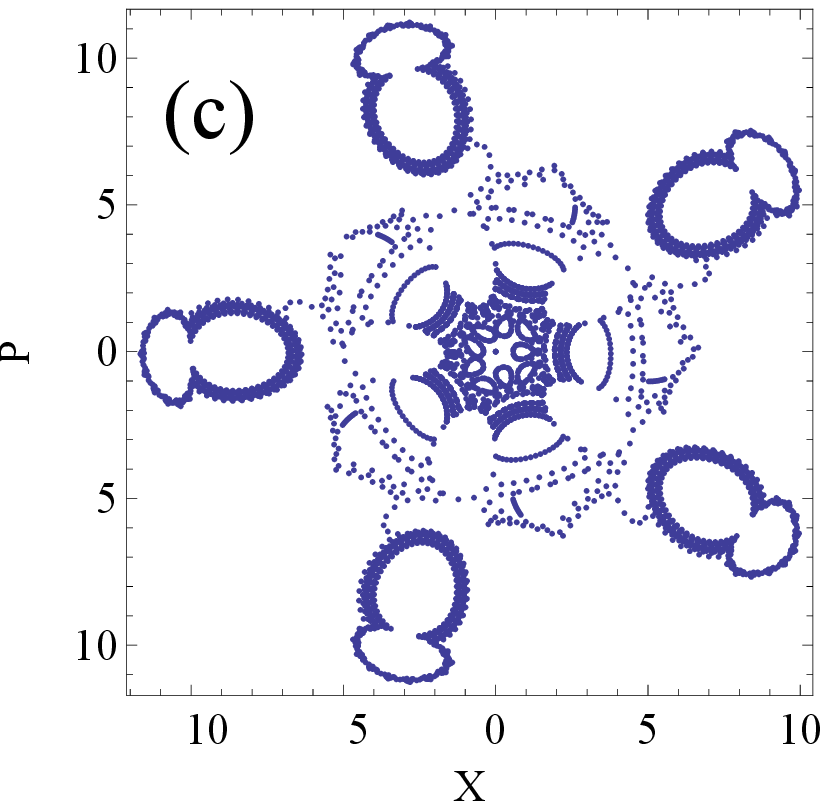}}
{\includegraphics[width=0.2\textwidth]{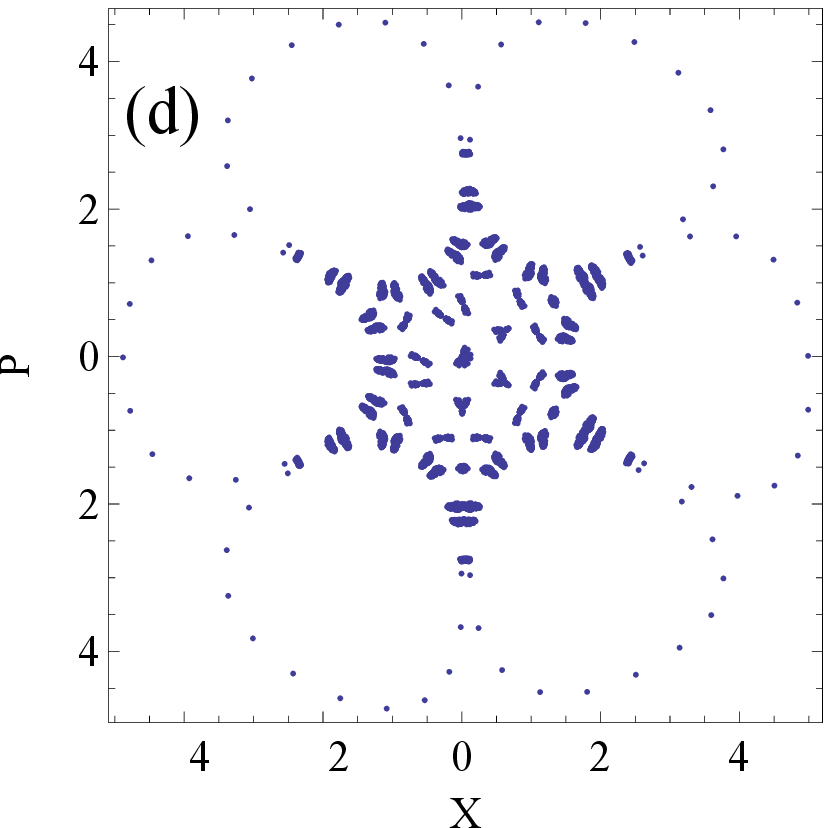}}
{\includegraphics[width=0.2\textwidth]{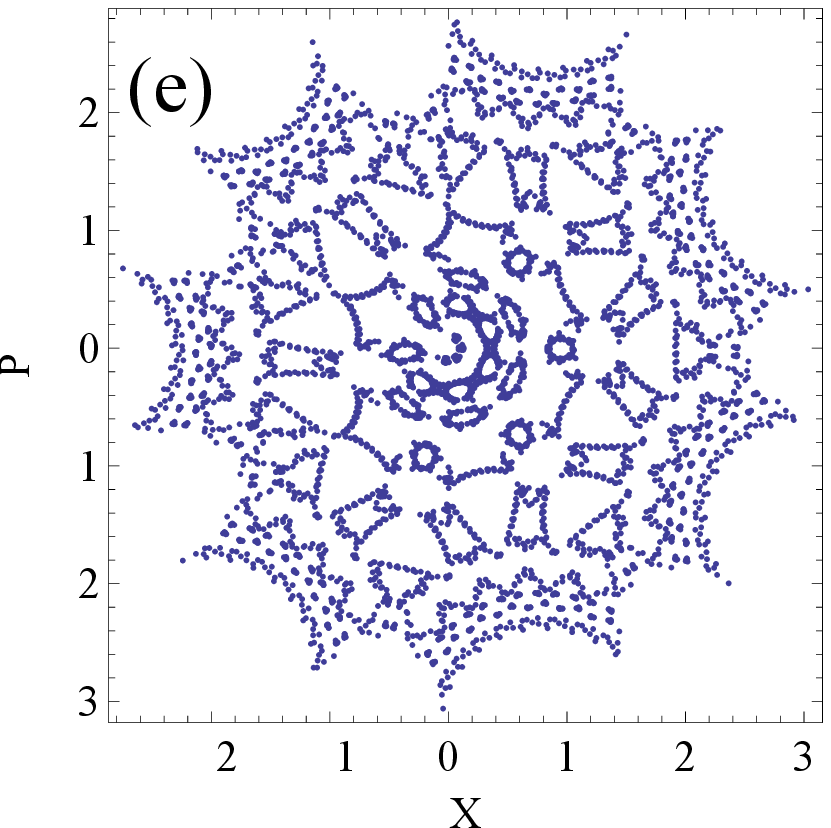}}
{\includegraphics[width=0.2\textwidth]{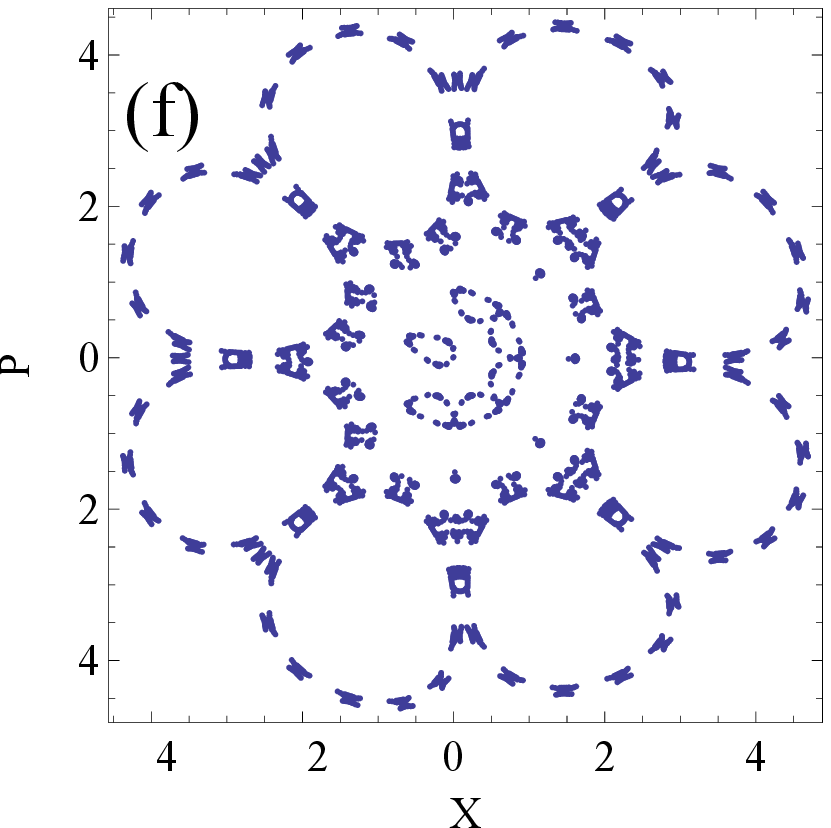}}
\caption{Phase space trajectories with just a single initial condition randomly chosen between (-0.5,0.5). We select the most appropriate phase space pictures in order to shed further light on the dynamics. All parameters are same as in Fig.~\ref{phase portraits}.} \label{random}
\end{figure}

Furthermore, the mapping described in Eq.~(\ref{web-map}) can be rewritten in more compact form as a single equation in terms of a single complex variable $z$, where $z_n=x_n + ip_n$, as
\begin{equation}
z_{n+1}= \left(z_n-iK\sinh \alpha x_n\right)e^{i\beta T} = \Theta(t) z_n, \label{map}
\end{equation}
where $\beta = \omega T$ and $z_n$ ($=x_n + ip_n$) is a complex number and its real and imaginary parts describe position and momentum of the moving membrane at the $nth$ impact. This more compact description is often advantageous when investigating symmetry properties of the system. For example, if $K=0$, the mapping will describe rotations through $\beta$ over one step. The evolution of the membrane in the presence of the radiation-pressure force, therefore, becomes
\begin{equation}
\Theta(t)=\left(1-\frac{iK\sinh \alpha x_n}{z_n}\right)e^{i\beta T}.
\label{tr}
\end{equation}
The function $\Theta(t)$, therefore, transforms the event $z_n$ of the phase space into $z_{n+1}$. It should be noted that, the transformation, as given in Eq.~(\ref{tr}), is a sum of two evolutions: The first term expresses the phase change over one period between short time interactions or kicks, and the second term expresses the influence of periodic kicks. In order to explore more about the stochastic web, after a few numerical experiments, we take single-trajectories of the several resonance cases starting from a random initial condition between $(-0.5,0.5)$ as shown in Fig.~\ref{random}.
Here, we select most relevant trajectories showing symmetrically stable and unstable regions of the stochastic web.
We also explore that the fixed points obey the relation
\begin{equation}
p_n=\left(\frac{1-a}{b}\right)x_n,
\end{equation}
which indicates that they appear along a straight line. The slope of the line depends on the value of $q$, which determines $a$ and $b$. Furthermore the line always passes through origin and produces $q$-fold symmetry.
\section{Stability Analysis Linearly}\label{sec:5}
We evaluate stability of our kicked membrane locally using eigenvalues based on straight-forward algebra. In order to start stability analysis linearly, the appropriate candidate is the Jacobian matrix written as,
\begin{equation}
\hat{J} = \left( \begin{matrix} a+bK \cosh x & b \\ -b+aK\cosh x&a  \end{matrix} \right).
\end{equation}
It can be seen that the determinant of the Jacobian is equal to one, that is $\vert \hat{J}\vert=1$, which implies that our web map is area preserving. In order to investigate the border between stable and unstable manifolds (trajectories), the eigen-values obey the relation $\lambda_\pm=\dfrac{1}{2}(T_r\pm \sqrt{T_r^2-4})$, where $T_r=2a+Kb\cosh x$ and $D_t=1$ are the trace and determinant of the Jacobian respectively. In order to examine the stability, these-eigen values directly lead us to study what is known as the exponential divergence in the evolution of the neighboring trajectories i.e. \textit{lyapunov exponent}.
\begin{figure}[H]
\includegraphics[width=0.45\textwidth]{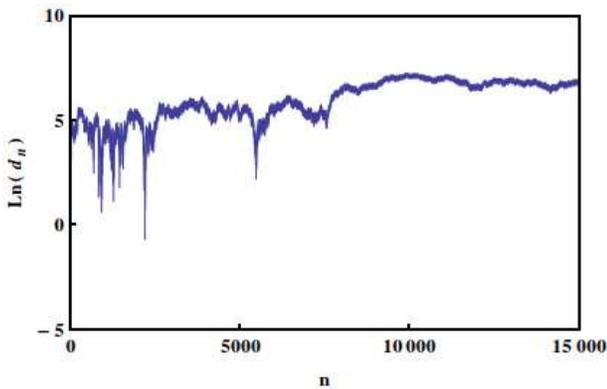}
\caption{Exponential divergence between two neighboring trajectories in a stochastic region of the phase space is shown via Lyapunov exponent for the case of $q=5$ and $K=0.5$.} \label{div}
\end{figure}
\begin{figure}[H]
\centering
\subfigure[]
{\includegraphics[width=0.23\textwidth]{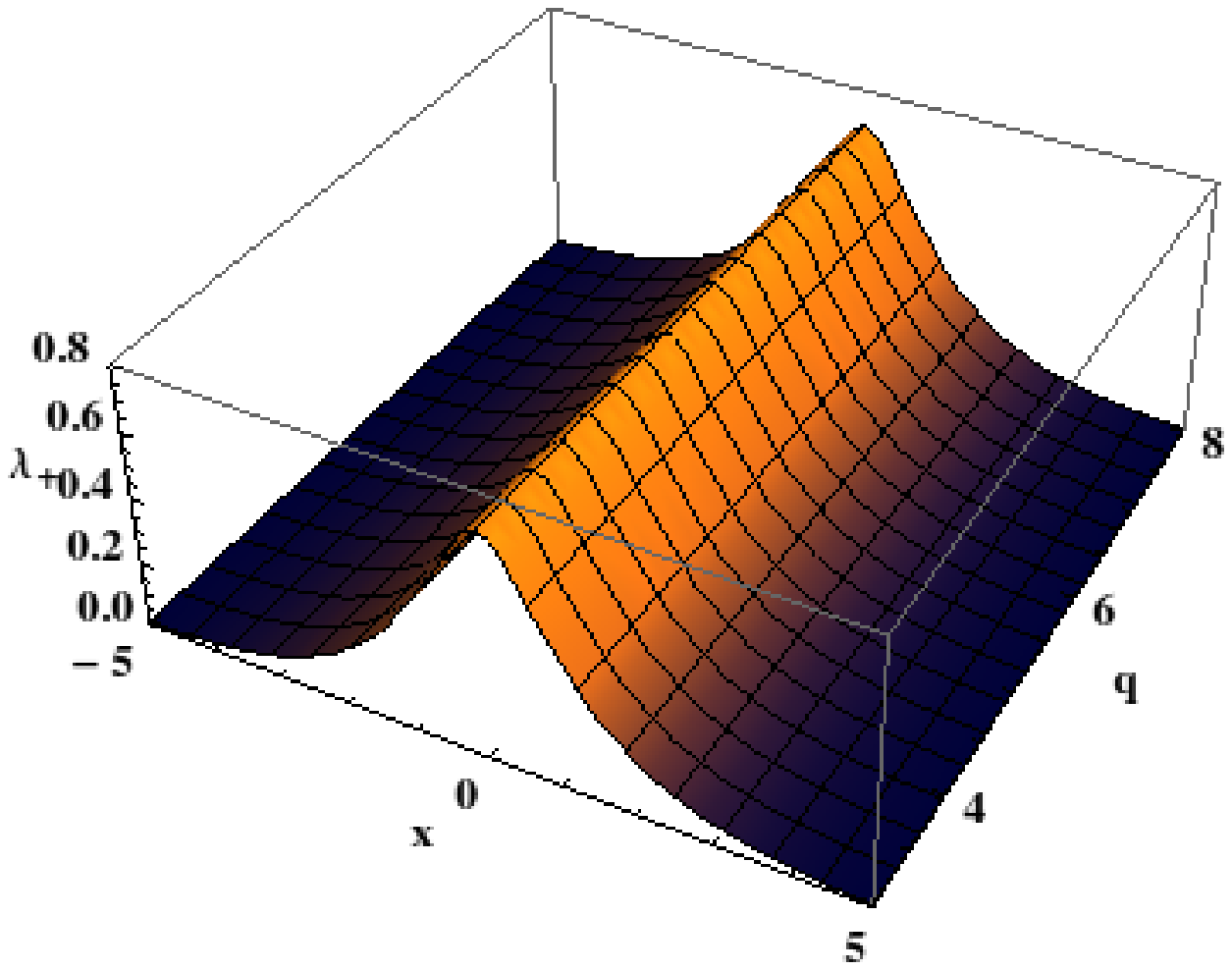}}
    \subfigure[]
{\includegraphics[width=0.23\textwidth]{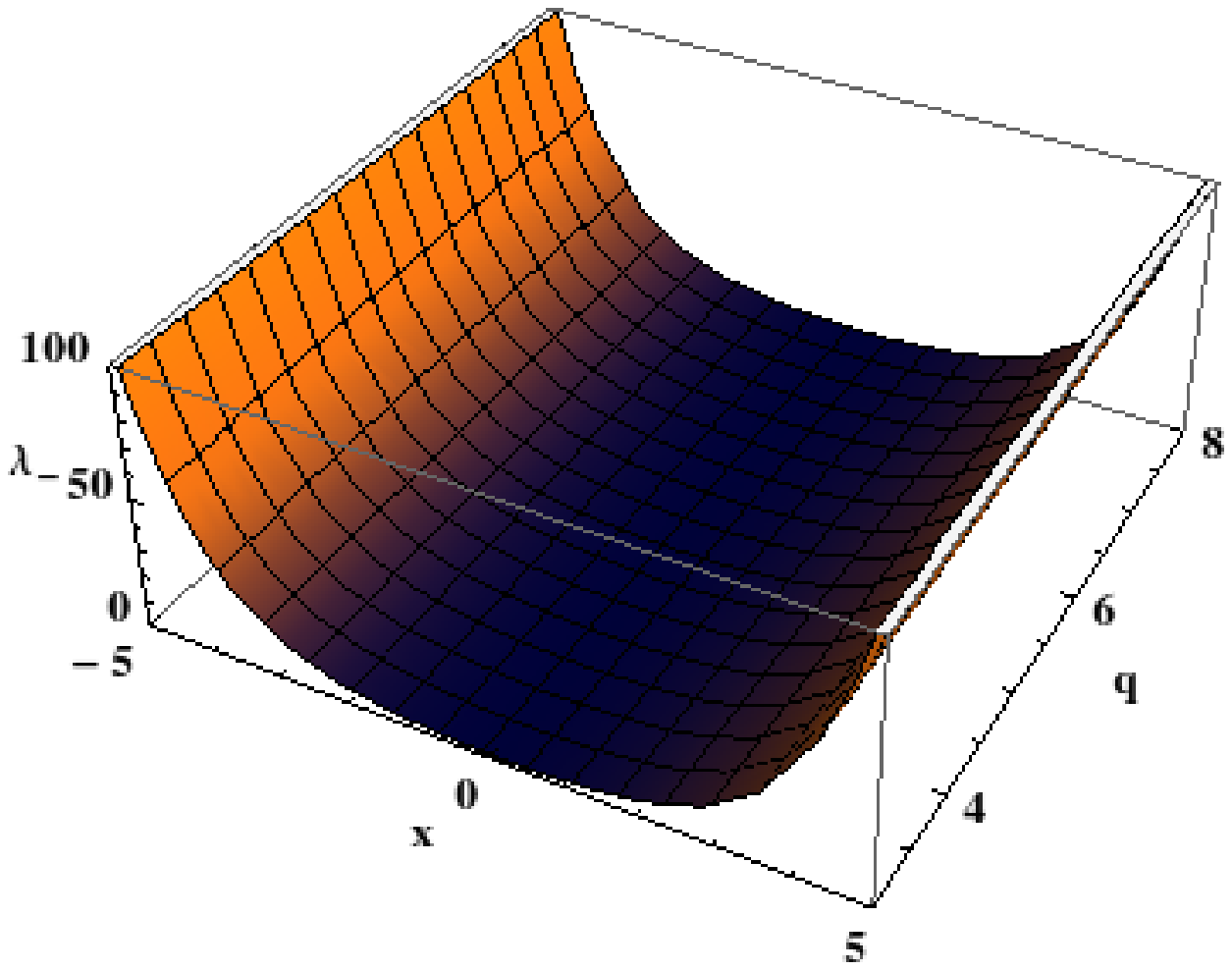}}
\caption{Here, we display $\lambda_\pm$, the eigen-values of the Jacobian matrix against $x$ and $q$.  The relation between $\lambda_\pm$ i.e. $\lambda_+$=$\frac{1}{\lambda_-}$ can be seen in (a) and (b). The entire phase plane is filled via similar growth rate under the pair of these eigen-values, therefore, forming the self-similar structures. Here $K=0.5$ whereas $q$ is varying from $q=3$ to $q=8$.}
\label{ev}
\end{figure}
If the initial coordinates are say $x_{01}$ and $x_{02}$, then at a later time they will have values $x_{01}(t)=x_{01}e^{\lambda_+t}$ and $x_{02}(t)=x_{02}e^{\lambda_-t}$. For our stochastic web map, eigen-values are complex for $-2<2a+Kb\cosh x<2$ and real outside of this interval. In the case of $q=3$ the evolution remains stable at origin, which is confirmed analytically as the corresponding real part is negative, whereas for all the rest of the cases the dynamics is unstable (Fig.~\ref{phase portraits}).

In order to describe exponential divergence (an important consequence of presence of chaos) of neighboring trajectories, we start with two different, but very close, initial points. Mathematically the Lyapunov exponent for the discrete map starting with some point $x_0$ is given by,
\begin{equation}
\lambda(x_0) = \lim_{n \to \infty} \dfrac{1}{n} \sum_{i=0}^{n-1} ln\mid f'(x_i)\mid. \label{LE}
\end{equation}
The distance between the iterates of these points should increase exponentially with time if the initial points are in a chaotic region of the phase space. A plot of the natural logarithm of this distance versus the number of iterations is shown in Fig.~\ref{div}. The above plot has been created using initial points $(15,0)$ which lies in a stochastic region and another in its neighbourhood viz., $(15.00001,0)$. Thus, the growth of lyapunov exponent indicates the presence of chaos in our system. Moreover, we plot the eigenvalues $\lambda_\pm$ against $x$ and $q$ in Fig.~\ref{ev}. These values,  $\lambda_\pm$ (either being positive in case of real or in case of complex the real part being positive) make the system unstable and positive lyapunov exponent indicates an exponential divergence.

Next we compute the survival probability of the trajectories of the phase portraits obtained in Fig.~\ref{phase portraits}. The survival probability is the probability for the oscillator to stay in a bounded region, $0<r<r_c$, and is given by \cite{28}
\begin{equation}
P_{s}(n) = 1-\lim_{N_{T} \to \infty} \dfrac{N_{E}}{N_{T}}.
\end{equation}
Here, $N_{T}$ is the total number of trajectories in the initial
ensemble and $N_E$ correspond to the number of trajectories that exited
the bounded region by discrete time (or number of kicks) $n$.
\begin{figure}[t]
\includegraphics[width=0.45\textwidth]{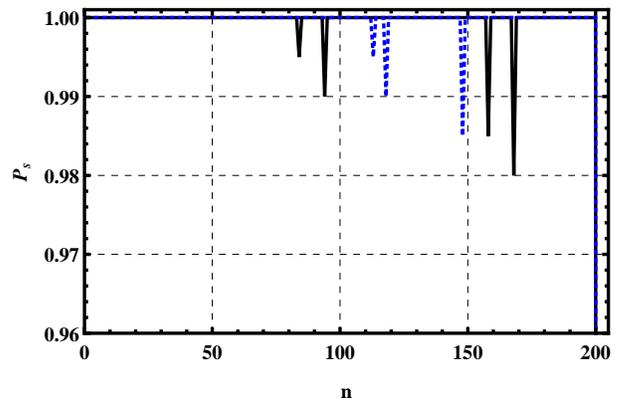}
\caption{The survival probability versus the separation time $n$ between the kicks is shown for the case of $q=5$. {\it Black solid-line} corresponds to $K=0.01$ and {\it blue dotted-line} corresponds to $K=0.1$.}\label{sp}
\end{figure}
As demonstrated in previous section, the stochastic web spreads the structures that fills the phase space with an infinite volume as long as $n \rightarrow \infty$. Physically, the initial conditions correspond to spreading trajectories all over the phase space. The survival probability $P_s$ provides us the information that what fraction of total trajectories survive in a particular region, whereas $1-P_s$ then corresponds to the information of escape of trajectories from that region. Another motivation behind $P_s$ comes from the fact that it corresponds exactly to the distribution of Poincar\'e surface of section that fills the entire phase space \cite{51}.
Fig.~\ref{sp} shows the survival probability for the case of $q=5$ for the different values of the kicking strength $K$. The (black) solid-line shows the survival probability for $K=0.01$ and (blue) dotted-line corresponds to the case of $K=0.1$. In our simulations, we consider the same initial point that we used for the Lyapunov exponent in Fig.~\ref{div} which lies in a stochastic region. It can be seen that, for higher value of the kicking strength $K$, more trajectories can survive in a particular stochastic region. 
\section{Conclusion}\label{sec:6}
In conclusion, we study the dynamical properties of mechanical nano- membrane in an optomechanical system. We show that the classical phase space of our system displays the presence of stochastic structures and $q$-\textit{fold symmetry}. It spans the entire phase space for the parameters $K$ and $q$. In order to understand the evolution of the system, in our analytical work we develop discrete web map. We find that the dynamics of the nano-mechanical membrane, in the presence of periodic delta kicks provided by radiation pressure force, displays quasiperiodic structures. We show that the change in the evolution takes place as the symmetry, resulting due to the condition on kicks at equal displacement, disappears. Furthermore, we explore that the q-fold symmetrical structures are compatible with chaoticity. It is shown that the Lyapunov exponent grows linearly showing the divergence of trajectories which is a confirmation of the presence of chaos in our system. Moreover, the values of parameters considered in our simulations satisfy current experimental values used in the laboratory experiments. For future perspective, the quantum studies of this system can be used for quantum information processing and quantum computing.
\section*{Acknowledgments}
We thank Higher Education Commission, Pakistan and Quaid-i-Azam University for financial support through Grants No. HEC/20-1374 and No. QAU-URF2014.

\end{document}